# Optical-pump terahertz-probe spectroscopy in high magnetic fields with kHz single-shot detection

Blake S. Dastrup*, Peter R. Miedaner*, Zhuquan Zhang, and Keith A. Nelson[‡]

*Department of Chemistry, Massachusetts Institute of Technology, Cambridge, MA, USA 01239*

*contributed equally to this work

[‡]corresponding author (kanelson@mit.edu)

**Abstract**

We demonstrate optical pump/THz probe (OPTP) spectroscopy with a variable external magnetic field (0-9 T) in which the time-dependent THz signal is measured by echelon-based single-shot detection at a 1 kHz repetition rate. The method reduces data acquisition times by more than an order of magnitude compared to conventional electro-optic sampling using a scanning delay stage. The approach illustrates the wide applicability of the single-shot measurement approach to nonequilibrium systems that are studied through OPTP spectroscopy, especially in cases where parameters such as magnetic field strength (B) or other experimental parameters are varied. We demonstrate the capabilities of our measurement by performing cyclotron resonance experiments in bulk silicon, where we observe B-field dependent carrier relaxation and distinct relaxation rates for different carrier types. We use a pair of economical linear array detectors to measure 500 time points on each shot, offering equivalent performance to camera-based detection with possibilities for higher repetition rates.

**Introduction**

Over the past few decades, there has been growing interest in using terahertz (THz) radiation as an ultrafast probe to study charge[1,2], lattice[3], and spin[4,5] degrees of freedom in a wide range of materials both in and out of equilibrium[6]. In some cases, THz time-domain spectroscopy has been combined with a high external magnetic field (B-field) to enable observations of B-field-dependent excitations and phases, such as electron cyclotron resonance (CR), electron spin resonance, and emergent quantum states[4,7–11]. An ongoing experimental challenge is to integrate this approach with additional pump sources to study photoexcitation dynamics. In a typical optical-pump THz-probe (OPTP) experiment, an optical pump pulse excites the sample into a transient non-equilibrium state, followed by a time-delayed THz probe pulse. Since the conventional THz detection method, electro-optic (EO) sampling, is also a time-resolved measurement in which the THz electric field (E-field) is probed with an optical gate pulse in an EO crystal and the relative delay is scanned in a stepwise manner, an OPTP experiment requires that at least two separate time delays be scanned to obtain a complete data set. As a result, this approach is experimentally time consuming, with full 2D scans often requiring whole days or more to achieve an adequate signal-to-noise ratio (SNR). This cost of time begins to become untenable when an additional scanning parameter, such as an external B-field, is introduced into the experiment.

One route to reduced data acquisition time is to replace stepwise EO sampling with echelon-based single-shot THz detection, in which a stair-step echelon is used to create a series of temporally and spatially separated beamlets that are overlapped with the THz probe in the EO crystal, and then imaged onto a camera so that each time step corresponds to a distinct set of pixels on the camera sensor[12–14]. Additionally, a recent technique relying on spectral encoding of temporal information was extended to work at high repetition rates with optical pumping[15]. However, while these methods have been successfully implemented and used for time-domain THz spectroscopy[16–18], including in pulsed magnetic fields[14], previous implementations have not enabled shot-to-shot balancing at kilohertz repetition rates due to the difficulty of retrieving THz signals at high speed using classical balanced detection[13,14]. One modification to the existing single-shot paradigm that was recently employed for two-dimensional THz spectroscopy uses a cylindrical lens to focus the two orthogonal polarizations of the gate beam onto a small region of the camera sensor to allow kHz readout[12]. Here we report the adaptation of this approach for the more widely conducted optical-pump THz-probe (OPTP) measurement, using a pair of economical 1D array detectors to perform shot-to-shot balanced detection in place of the specialized and expensive high-speed camera used previously. To test the capabilities of our setup, we measured the spectro-temporal evolution and B-field dependence of electron CR in bulk high-resistivity Si following an 800 nm optical excitation pulse. The results illustrate versatile applicability of the method to a wide range of OPTP measurements in which optical pumping gives rise to electronic, vibrational, or spin evolution that can be probed by THz spectroscopy.

**Experimental**

A schematic illustration of our high B-field single-shot OPTP setup is shown in Fig. 1. The setup consists of two high energy pump beam paths and one low energy gate beam path all derived from the 4 mJ, 100 fs output of a Ti:Sapphire regenerative amplifier (Coherent Astrella F-1K) with an 800 nm center wavelength. The gate beam used for single-shot readout of the THz signal is split from the main beam with a 90:10 beam splitter. The remaining portion of the main beam is split using a 60:40 beam splitter into the two pump beams with energies of roughly 2 mJ and 1.6 mJ used for THz generation and optical pumping respectively. The THz generation beam is directed onto a 1 mm thick zinc telluride (ZnTe) crystal with a 10 mm diameter (Ø) clear aperture (Eksma optics), which is completely filled by the 11 mm spot size of the THz generation beam. A slowly diverging THz beam (Rayleigh range ~ 0.25 m at 1 THz) emerges from the crystal and propagates into the Ø 100 mm bore of a superconducting solenoid magnet (0-9 T) parallel to the applied magnetic field, $\boldsymbol{B_{ext}}$. A four-rod aluminum cage structure extends through the magnet bore (500 mm length) and is mounted to the optical table at both ends of the magnet. A pair of 1-inch paraboloid mirrors (90°, EFL = 1") are fixed to an optical mount that is mounted to the four cage rods and positioned at the B-field center (see Fig. 1b). The THz beam is focused into the sample by one of the paraboloids, and the transmitted THz signal is collected by the other and directed out of the bore parallel to $\boldsymbol{B_{ext}}$. Outside the magnet bore, the THz beam is collected by a 2-inch paraboloid mirror (15°, EFL = 20") and then focused by a second 2-inch paraboloid mirror (90°, EFL = 4") onto a 2 mm thick ZnTe crystal for EO detection. The 1.6 mJ beam used for optical pumping is delayed with an optical delay line and then demagnified with a 3:1 telescope (L1′:L2′) to a spot size of ~3 mm. A periscope is used to displace the optical pump beam vertically with respect to the THz generation beam, after which the beam is directed into the magnet bore parallel to the THz generation beam. In the bore a Ø 0.5" silver mirror mounted to the cage mount above the focusing paraboloid reflects the optical pump onto the sample at a polar angle of approximately 46° with respect to the sample normal.

The EO sampling gate beam is delayed using an optical delay line and then expanded by a total of 9x with two separate 1:3 telescopes (L1:L2 and L3:L4) to create a roughly uniform intensity profile that fills the echelon. The stair-step echelon (Sodick) is a silver-coated reflective optic with 500 steps of 60 µm width and 7 µm height. The enlarged beam is reflected from the echelon, creating a sequence of 500 pulses with an interpulse delay of ~40 fs, which corresponds to an overall time window of ~20 ps. The beam at the echelon plane is relayed to an image plane intermediate between the echelon and detection crystal using a 4f imaging system (L5:L6). The image is then focused by L7 and recombined with the THz beam using a pellicle beamsplitter (PBS) so the two beams are focused together onto the EO detection crystal. After the crystal, the optical beam is recollimated by lens L8, which forms the end of a second 4f imaging system (L7:L8). A third 4f imaging system (L9:L10) relays the image after L8 to the detection plane. Inserted into the beam path are a quarter waveplate, which serves to balance the intensity of the gate beam into any set of orthogonal polarizations, and a 10° Wollaston prism, which separates the gate beam into two orthogonal polarizations. The two polarization components are parallelized and focused by a compound cylindrical lens (a pair of two separated cylindrical lenses, shown in Fig. 1c) set to achieve both focusing and optimal spacing (see SI) onto a pair of parallel 1D detector arrays (Synertronic Glaz LineScan-I-Gen2) with 12 mm spacing, which are read out at a 1 kHz frame rate to achieve shot-to-shot balancing.

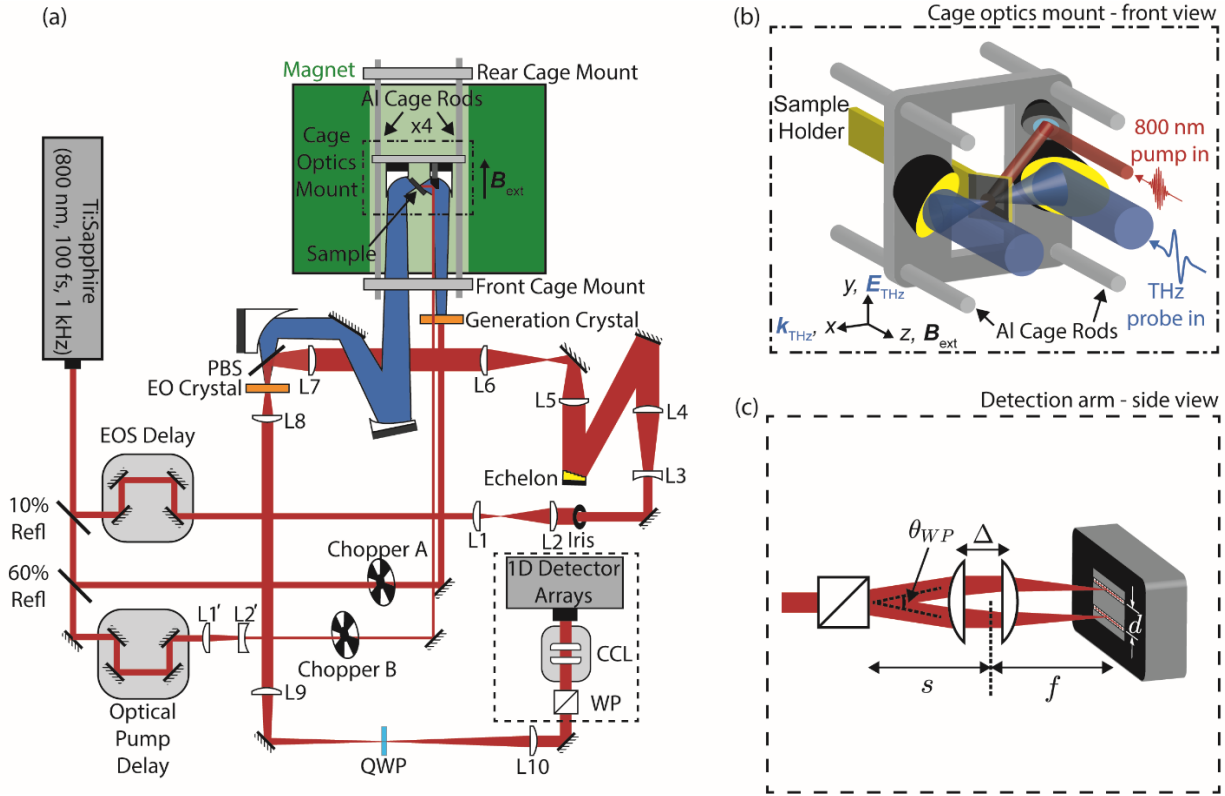

**Figure 1.** Single-shot high B-field OPTP optical setup. (a) Schematic of experimental setup. $B_{ext}$: external B-field; BS: beam-splitter; $k_{THz}$: THz propagation direction; $E_{THz}$: THz E-field polarization direction; HWP: half waveplate; QWP: quarter waveplate; WP: Wollaston prism; CCL: compound cylindrical lens; L1-L4, probe beam expanding telescopes (1:3); L5-10, three successive 4f imaging systems ($f$ = 10, 10, 10, 5, 30, 30 cm); L1' ($f$ = 20 cm); L2' ($f$ = -7.5 cm); PBS: pellicle beam-splitter. (b) Front view of in-bore cage optics mount. THz beam path shown in blue. Optical pump path shown in red. Sample mount with 45° tilt depicted in gold. (b) Side view of probe beam path through Wollaston prism and compound cylindrical lens which focuses the two probe arms onto the two 1D array detectors.

We employ a differential chopping data collection scheme using two choppers that are inserted into the two pump beam paths and triggered using the 1 kHz trigger obtained from the laser. Labeling the THz generation beam as "A" and the optical pump as "B", chopper A runs at 500 Hz blocking every other pulse, while chopper B runs at 250 Hz blocking every other set of two pulses. Together, the two choppers produce each of the four combinations $A_{on}/B_{on}$, $A_{on}/B_{off}$, $A_{off}/B_{on}$, and $A_{off}/B_{off}$ over the course of 4 consecutive laser shots. We can thereby obtain the THz probe waveform (A), optical pump reference (B), the THz-probe waveform with optical pumping (AB), and the differential signal as follows,

$$\frac{\Delta I_\alpha}{I} = \frac{I_\alpha^+}{2I_0^+} - \frac{I_\alpha^-}{2I_0^-} = \frac{I_\alpha^+ - I_\alpha^-}{I_0} \qquad (1)$$

$$\frac{\Delta I_{\text{diff}}}{I} = \frac{\Delta I_{AB}}{I} - \frac{\Delta I_A}{I} - \frac{\Delta I_B}{I} \qquad (2)$$

where $I_\alpha^+$ and $I_\alpha^-$ are the raw 1D gate arrays for the two polarization arms of the balanced detection scheme ($\alpha = \{A, B, AB\}$), and $I_0^+$ and $I_0^-$ are the reference 1D gate arrays obtained when both pump beams are blocked by their respective choppers. The final equality in Eq (1) is true only when $I_0^+ = I_0^- = I_0/2$, which implies perfect balancing.

We observe slight deviations (no larger than ~5%) in balancing from pixel to pixel, which we estimate leads to uncertainties in THz field amplitudes of ~0.5% (details in SI).

THz transmission measurements were done in the Voigt geometry ($k_{THz} \perp B_{ext}$). A [100] undoped high-resistivity Si sample with 0.5 mm thickness was mounted in a liquid helium cryostat with c-cut sapphire windows and a 1" path length from window to window. The cryostat was fixed to the mounting plate of a 600 mm linear stage and inserted into the gap between the two 1" paraboloid mirrors through a rectangular opening in the mount. For THz transmission experiments, the sample was oriented with the [100] axis either at 90° or 45° with respect to $B_{ext}$ (where the angle is a rotation in the [011] plane of the sample). Measurements in the 90° orientation were made by mounting the sample to a copper sample mount with a Ø 2 mm aperture, while measurements in the 45° orientation were made by mounting the sample to an angled copper sample mount with Ø 3 mm clear aperture. In both cases the sample was cooled to a temperature of 5 K. The 800 nm optical pump was set to a fluence of ~0.1 mJ cm$^{-2}$. Using higher pump fluence led to the loss of CR signal.

**Results and Discussion**

To test the performance of our single-shot detection setup, we compare the THz waveform and spectrum measured by single-shot detection to that measured by conventional detection (see Figs. 2a and 2b). The THz signal was measured in the absence of a sample by both methods using the same optical setup (for conventional detection two additional mirrors were inserted into the gate beam path to bypass the echelon, and photodiodes were used in place of the 1D detector arrays). Fig. 2c shows the raw 1D arrays obtained from each 1D array averaged over 1000 shots. The THz waveform shown in Fig. 2b was extracted from the raw signal by calculating $\Delta I_\alpha/I$, as given in Eq. 1, on a shot-to-shot basis and averaging over 10,000 shots (single-shot) or 500 shots per time point (conventional). For single-shot detection, this extraction yields the THz waveform as a function of pixel number. We then used the calibration procedure described by Gao et al.[12] to correlate pixel number with the real time axis as well as to correct for amplitude non-uniformity.

A comparison of the noise levels obtained by single-shot detection versus conventional detection was made by plotting the noise level of a 20 ps time window against the number of total laser shots collected for a given measurement (see Fig. 2d). The gate delay line was positioned to detect a featureless time window before the arrival of the THz pulse.

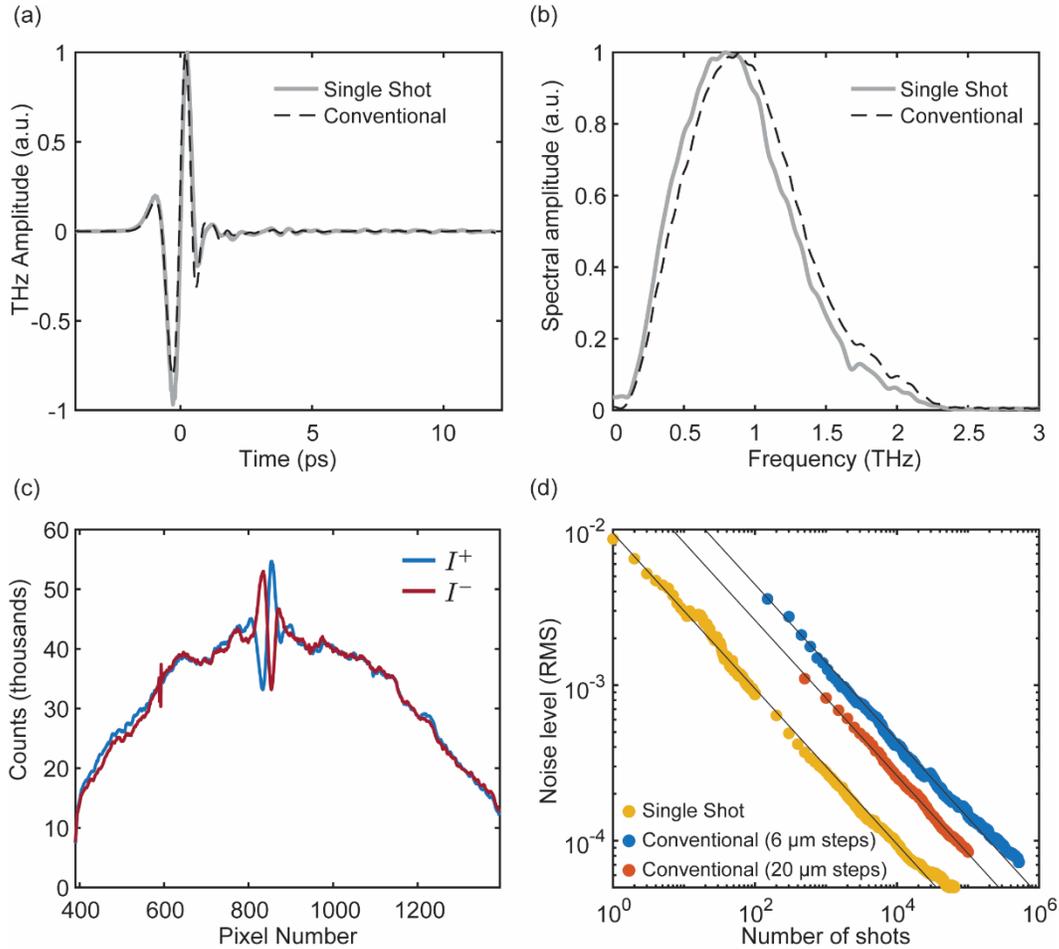

**Figure 2.** Characterization of single-shot detection performance. (a) Time-domain and (b) frequency-domain comparison of THz waveform collected with single-shot detection (solid gray line) and conventional stepwise detection with 6 μm steps (dotted black curve). (c) Raw probe signal on 1D detector arrays with THz signal present. The plus and minus polarizations are normalized by the probe reference (without THz signal) and then subtracted from each other on a shot-to-shot basis. (d) Noise level analysis comparing the rms noise level in a featureless region of the time-domain signal for single-shot (yellow), conventional detection with 20 μm time-steps (red), and conventional detection with 6 μm time-steps (blue).

Each single-shot measurement captures a 20 ps time window and contains 500 time points, so a conventional scan with an equivalent step width requires 500x the number of laser shots to achieve the same time window. We performed two sets of conventional scans with 6 μm (40 fs) step width and 20 μm (133 fs) step width, the latter to allow for a more direct comparison of noise levels for measurements that don't require the spectral bandwidth (~10 THz) afforded by the short time step. Our measurements show more than an order of magnitude lower noise level for single-shot detection over conventional detection for an equal number of shots, which is consistent with the previous report utilizing a high-speed camera as the detector[12].

To showcase the measurement capabilities of our setup, we have measured electron CR in bulk undoped Si following optical excitation at 800 nm in an external B-field which was swept from 0-9 T. THz-frequency CR in bulk undoped Si has previously been reported for a single-shot THz measurement in a pulsed magnetic field at low repetition rate[14]. Those measurements were conducted with a fixed (100 ps) delay between the optical pump and THz probe pulses. Time-resolved carrier dynamics have been measured in other systems by techniques such as optical reflectivity[19] and far-IR transmission spectroscopy[20,21]. Our measurement scheme allows for the collection of the full spectral content of the THz probe with the delay $\tau$ between the optical pump and THz probe varying over a 600 ps range and the external B-field varying from 0-9 T, which provides additional spectral information that allows the dynamics of different carrier types to be observed independently. Raw time-domain plots and time traces at selected optical-pump/THz-probe time delays are shown in Fig. 3. The acquisition time of the complete B-field sweep (0-9 T) was 10.3 hrs. We estimate that the present study would have required more than 11 days of data acquisition with conventional THz detection. The dependence of the CR signals on the sampling delay time t is discussed briefly in Supplementary Information section S9. Figure 4 shows plots of the THz absorption spectrum as a function of optical-pump/THz-probe delay at different B-fields. The absorption spectrum, $A(\Omega)$, was calculated as,

$$A(\Omega) = -\log\left(\frac{\Delta I_{AB}/I}{\Delta I_A/I}\right) \tag{3}$$

which is valid in the linear response regime. The plots show a zero-frequency Drude response at 0 T that decays within ~200 ps. As the B-field is increased, this zero-frequency response resolves into a set of four absorption peaks that undergo a blue-shift with increasing B-field. These four resonances can be attributed to two electron and two hole carrier types with distinct effective masses that have been observed in previous CR experiments in bulk Si[22]. Charge carriers are generated when absorption of the 800 nm pump promotes electrons into any of the six equivalent conduction band valleys near the X-point at the zone boundary, leaving holes in the two valence bands at the Si Γ-point. Under the parabolic band approximation, the constant energy surfaces of the valence band maxima are spheroidal, and those of the conduction band valleys are ellipsoidal[22]. These ellipsoids have principal axes oriented along the crystalline axes such that when the sample is rotated, the curvature of the path traversed by the electron in *k*-space is altered, leading to the observation of an orientation-dependent effective mass[22,23]. Because the applied B-field is oriented in the [011] plane of the sample, the CR frequencies from conduction band electrons in the four valleys in the [100] plane will be equivalent, as well as the frequencies of electrons in the two valleys parallel to the [100] axis (in the [011] plane). We thus refer to the electron resonances as [100] electrons from the four valleys in the [100] plane, and [011] electrons from the two remaining valleys in the [011] plane. We refer to the two hole resonances as light and heavy holes. From the absorption spectra we extracted the frequency of each resonance via simultaneous fitting to four Lorentzian functions. In Fig. 4b&d we have plotted the center frequencies of the four resonances obtained from these fits as a function of the applied B-field. We have excluded points at low B-field where the resonances are not sufficiently resolved to give good fits, as well as points at $B = 8, 9$ T for the light hole in the 45° sample orientation where the resonance is near the edge of the THz bandwidth and a good quality Lorentzian fit could not be obtained. In the 45° sample orientation, the [100] and [011] electron CR frequencies are nearly degenerate, such that the peaks in the THz absorption spectrum corresponding to these resonances have a high degree of overlap at low B-field. The frequency of the light hole lies just above those of the two electrons and the heavy hole just below. The corresponding effective masses for these resonances in order of increasing mass are $m^* = \{0.19 \pm 0.003, 0.26 \pm 0.003, 0.30 \pm 0.005, 0.50 \pm 0.02\} m_e$ which are in good agreement within our experimental uncertainty with values reported by Dresselhaus et al[22] ($m_e$ is the free electron mass). For the 90° sample orientation, we observe the corresponding effective masses $m^* = \{0.19 \pm 0.02, 0.244 \pm 0.002, 0.42 \pm 0.03, 0.64 \pm 0.03\} m_e$, which agree within uncertainties with the literature values but note that the heavy hole mass deviates from the expected trend in which the mass decreases going from 45° to 90° (see Fig. S7).

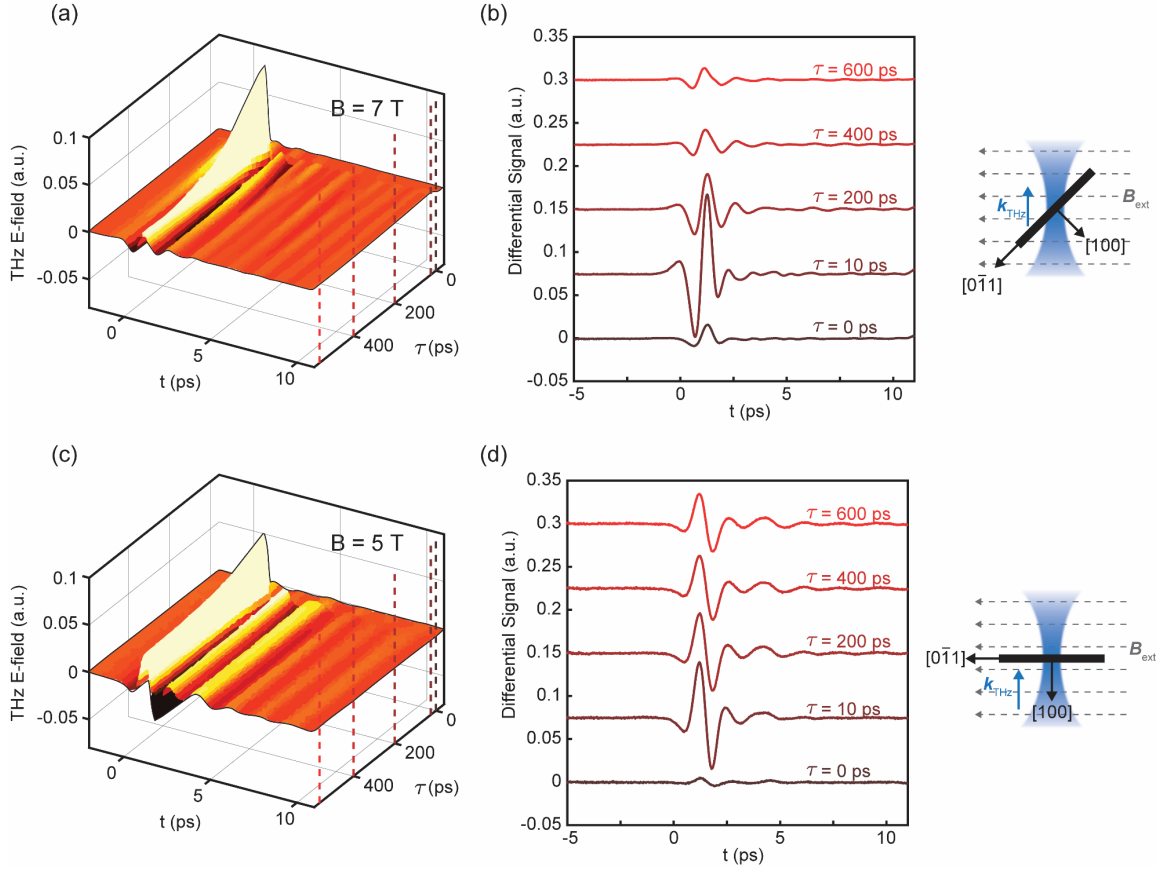

**Figure 3.** Representative cyclotron resonance time-domain data from [100] high-resistivity silicon. Data were collected with $E_{THz} \perp B_{ext}$ and $k_{THz} \perp B_{ext}$ (Voigt geometry). 2D time-time plots are shown for sample orientations of (a) 45° and (c) 90° with respect to $B_{ext}$ at B = 7 T and B = 5 T respectively with optical-pump/THz-probe time delays $\tau = \{0, 10, 200, 400, 600\}$ ps. Constant $\tau$-slices are plotted for the same sample configurations and B-fields in (b) and (d). The measurement configuration for each set of plots is shown by the diagram on the right.

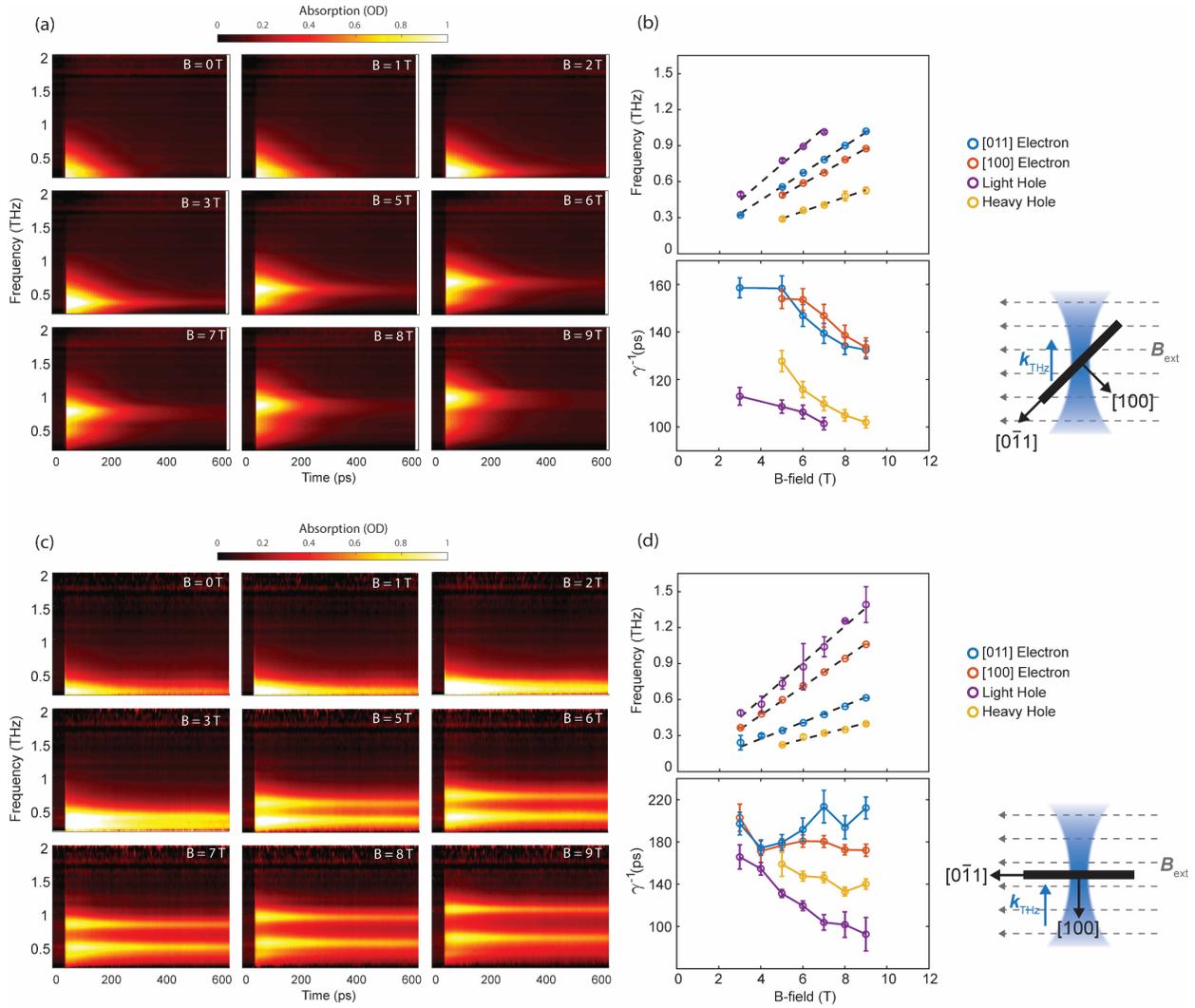

**Figure 4.** Cyclotron resonance spectra from [100] high-resistivity silicon. THz absorption as a function of optical pump/THz time delay plotted for $|B_{ext}| = $ 0-9 T and samples oriented at (a) 45° and (c) 90° with respect to $B_{ext}$ as depicted by the diagrams on the right hand side of the figure. (b),(d) CR center frequencies (top) and inverse decay constant (bottom) as a function of external B-field. Error bars are the 95% confidence intervals extraced from (top) Lorentzian fits of the resonances and (bottom) exponential fits of the decay curves. The relationship between CR frequency and carrier effective mass is given by $\omega_c = eB/m^*$, where $e$ is the electron charge, and $m^*$ is the effective mass. We see four distinct resonances at both 45° and 90° sample orientation. Carrier relaxation is resolved into the four carrier types and we see distinct decay times for each carrier as well as a general trend of decreasing carrier lifetime with increasing B-field, with some exceptions.

By measuring the entire 2D data set, we get complete spectral and temporal information about the photoinduced THz absorption. For example, we see slightly different decay behaviors for the different carrier types as shown in the bottom panels of Figs. 3b,d. To obtain these plots, decay curves for each of the four resonances were fit to an exponential function of the form, $f(\tau) = \exp(-\gamma\tau)$, where $\gamma$ is the decay rate constant. We observe that the electrons decay more slowly than the holes for both the 45° and 90° sample orientations. We also observe that all four of the carrier types exhibit B-field-dependent decay times. For the 45° sample orientation, the decay time constant for all four resonances decreases nearly monotonically with increasing B-field. For the 90° sample orientation, we also see clear differences among decay time constants for the different carrier types as well as B-field dependence. In this case, the decay time of the hole resonances decreases with increasing B-field, while that of the [011] electron is relatively flat, and the decay time of the [100] electron appears to increase with increasing B-field. B-field dependence of carrier relaxation has previously been observed in graphene[19,24], where it was argued that the nonuniform spacing of Landau levels in the presence of a B-field gave rise to a decrease in Auger scattering events, and thereby an increased carrier lifetime. Here we generally observe the opposite trend (relaxation rate increases with increasing B-field). However, Landau level splitting in bulk Si is essentially harmonic, giving rise to states are symmetrically distributed about the carrier energy. Furthermore, Landau level degeneracy increases with increasing B-field[25], which leads to an increased density of states in the vicinity of the carrier energy. These two effects together make it likely that the changes in level spacing and degeneracy alter the available scattering pathways as the magnetic field increases, which, in turn, leads to an observable increase in the decay rate of excited carriers.

**Conclusion**

We have constructed a variable B-field optical-pump/THz-probe spectroscopy setup with single-shot detection that allows for shot-to-shot balancing at a 1 kHz repetition rate with the use of 1D linear array detectors. We observe SNR on par with what has recently been demonstrated for single-shot 2D THz spectroscopy based on a high-speed camera. This setup allows for the collection of full 2D optical-pump/THz-probe scans at different magnetic fields, effectively a 3D measurement, in a dramatically reduced amount of time compared to conventional EO sampling of the THz probe. The present cyclotron resonance measurements of photoexcited carriers in silicon have yielded magnetic field-dependent electron and hole effective masses and decay rates that are distinct among the different carrier types. The capabilities demonstrated in this measurement have the potential to enable new dynamical studies of low-energy magnetic excitations and other phenomena in a wide range of materials. The single-shot measurement approach will facilitate systematic optical-pump/THz-probe studies with any of a wide variety of variables including optical and/or THz pulse fluence or polarization, applied electric field or pressure, and others.

**Acknowledgements**

This work was supported by the U.S. Department of Energy, Office of Basic Energy Sciences, under Award No. DE-SC0019126

# Supplementary Information for optical-pump terahertz-probe spectroscopy in high magnetic fields with kHz single-shot detection

Blake S. Dastrup*, Peter R. Miedaner *, Zhuquan Zhang, and Keith A. Nelson

*Department of Chemistry, Massachusetts Institute of Technology, Cambridge, MA 02139*

## S1. *Probe detection alignment*

We used a compound cylindrical lens design—two cylindrical lenses with adjustable spacing—to achieve simultaneous focusing of the two probe polarizations with correct beam spacing at the detection plane of the 1D detector arrays. The compound lens focal length for two identical ideal lenses is given by,

$$\frac{1}{f} = \frac{2}{\tilde{f}} - \frac{\Delta}{\tilde{f}^2} \qquad \text{(S1)}$$

where $f$ is the compound lens focal length, $\tilde{f}$ is the focal length of the individual lenses, and $\Delta$ is the lens spacing. By requiring the distance from the compound lens to the detector plane to be $f$ and following a simple ray optics analysis of the detection arm of the probe (see Fig. 1b) one can derive the relation,

$$f = \frac{d}{\tan(\theta_{\text{WP}}/2)} \qquad \text{(S2)}$$

where $d$ is the beam spacing, and $\theta_{\text{WP}}$ is separation angle of the Wollaston prism (8° in our case). This equation specifies the lens focal length needed to achieve the desired beam spacing. Note that the beam spacing, $d$, does not depend on the distance, $s_1$, from WP to the lens. Consequently, alignment of the probe can be performed without the need to account for changes in this distance. Our procedure for aligning the probe beams onto the 1D detector was to make a small adjustment of the beam spacing by adjusting the lens spacing to change the focal length, then adjusting the position of the compound lens so that the principal plane is focused at the detector plane, then repeating. In this way, the optimal lens focal length can be found by an iterative optimization process.

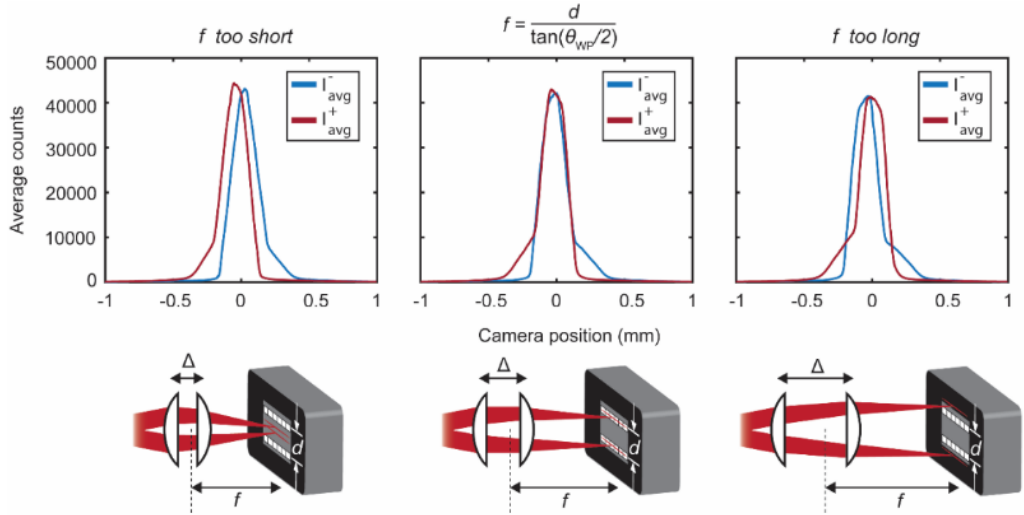

**Figure S1. Procedure for optimizing compound lens focal length.** The three plots above show data collected by scanning the line detector pair vertically and plotting the pixel averaged probe signal vs. vertical detector position. When the two probe polarizations are each centered on their respective line detectors, the pixel-averaged signal will show maximal overlap, as in the center plot. If the focal length of the lens pair is too long or too short, the pixel-averaged signals will peak at different vertical displacements as in the right-hand and left-hand plots. Beneath each plot is an illustration of the corresponding lens spacing.

The 1D detector arrays were mounted on a vertical linear stage with a motorized actuator to enable fine motorized control of the vertical position of the detectors. We developed a procedure for monitoring the probe alignment process after each iteration and setting the vertical position of the detectors, which involved scanning the vertical position of the detectors and measuring the probe signal on both detectors as a function of camera position. Fig. S1 shows camera alignment traces. When the lens spacing is optimized the signals from the two probe arms overlap completely, whereas deviations from the optimal lens spacing cause the signals to be offset.

## S2. *Camera positioning in B-field*

We observe that in the presence of a magnetic field, the optical alignment in our setup changes slightly as evidenced by small changes in probe balancing (and therefore THz signal amplitude), which we attribute to B-field induced torques experienced by optical elements in close proximity to the magnet. In particular, the THz signal amplitude decreases with increasing B-field, but no sharp frequency-dependent changes are observed in the THz spectrum with increasing B-field. In order to compensate for these alignment changes we employ an automatic rebalancing procedure after each change in the external B-field by vertically scanning the camera position, and calculating the balancing deviation function, $g(z)$, given by,

$$z^* \equiv \underset{z}{\mathrm{argmin}}\{g(z)\} = \underset{z}{\mathrm{argmin}}\left\{\frac{1}{N}\sum_{n=1}^{N}\left|1 - \frac{I_n^+}{I_n^-}\right|\right\} \tag{S3}$$

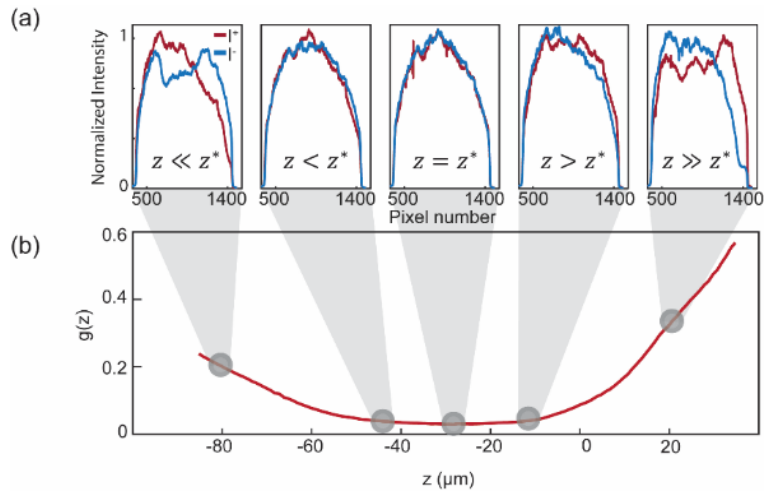

**Figure S2. Procedure for optimizing vertical position of detector** (a) Examples of raw signal from detector positioning procedure. Vertical positioning, $z$, is swept and the quality of balancing reaches a maximum, and then diverges. (b) Example of objective trace for 5 T field. Note that the minimum of the function occurs at a position shifted slightly below zero.

where $n$ is the pixel number, and $z^*$ is defined to be the $z$-position where $g(z)$ is minimized. Once the scan is complete, the camera position is set to $z^*$.

## S3. *Relaxation fits and decay time extraction*

Decay time constants were extracted from exponential fits of the decay curves at each CR frequency. The decay curves were obtained by plotting the amplitude of the absorption spectrum at the resonance center frequency as a function of pump-probe delay time, $\tau$. To systematically determine the center frequency of the four resonances, the THz absorption spectrum at $\tau = 200$ ps was fit to four Lorentzian functions. The exponential fits are shown in Fig S3 and Fig S4. As mentioned in the main text, decays below 2T are excluded because of the inability to resolve individual resonances at low field. Heavy hole data are excluded below 4T for the same reason, below this field strength the resonance cannot be properly resolved. Light hole data in the 45° sample orientation are excluded because the signal level is low at this frequency leading to poor fits at long pump probe delays.

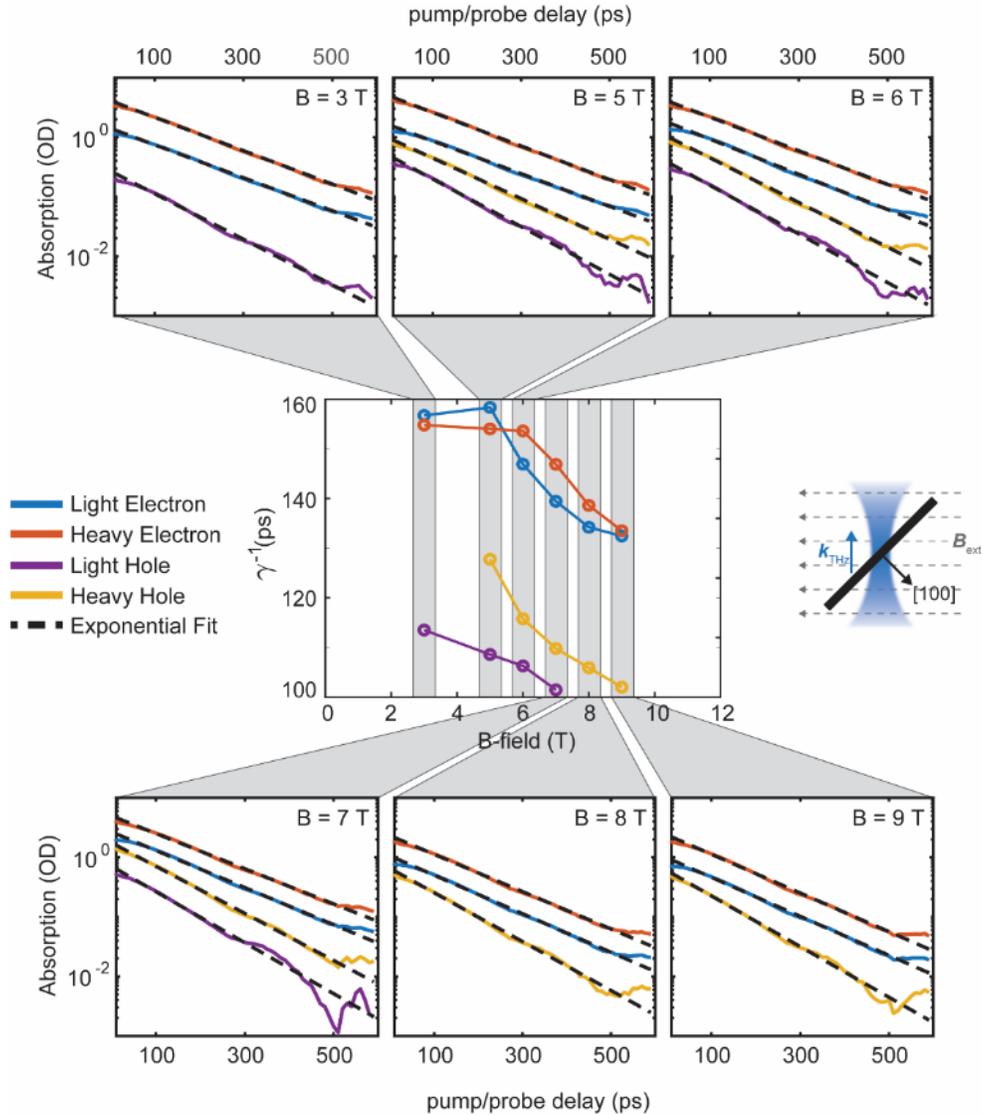

**Figure S3. Exponential fits of carrier decay at 45° sample orientation.** Decay curves for the different carrier types are plotted on a log scale in the panels above and below the center plot. The carrier types are the light electron (blue), heavy electron (orange), light hole (purple), and heavy hole (yellow) resonances with fits shown as dashed black lines. The fit lines were obtained by fitting the experimental decays to an exponential function of the form $f(\tau) = Ae^{-\gamma\tau}$, where $\gamma^{-1}$ is the carrier lifetime. The carrier lifetime is extracted for each carrier type at each B-field and plotted as a function of B-field in the center panel, which is also shown in Fig. 4 of the main text. The data shown here are for the 45° sample orientation as shown by the diagram to the right of the center plot. Curves for different carrier types have been vertically offset from each other for clarity.

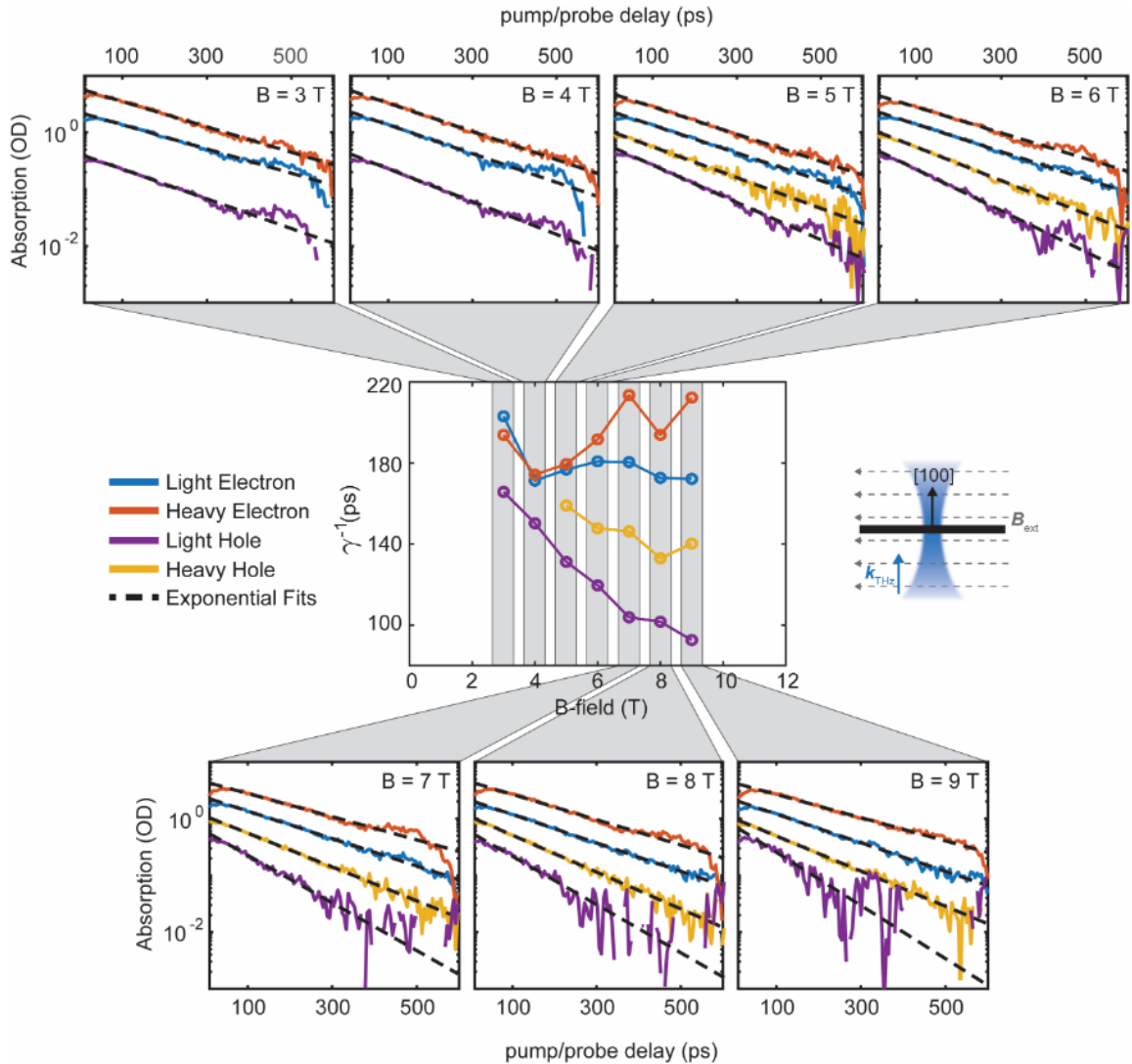

**Figure S4. Exponential fits of carrier decay at 90° sample orientation.** See caption for Fig. S3.

## S4. *Time-domain data*

Implementing a rapid data collection scheme allows for full dual-delay datasets to be collected in a reasonable amount of time and allows for the adjustment of other experimental variables such as magnetic field. Shown here are the full datasets collected along two separate time axes (EO sampling delay, $t$, and optical-pump/THz-probe delay, $\tau$). Each of the full B-field scans took ~10 hours to acquire. We estimate that to collect an equivalent data set with the same S/N level using conventional EO sampling would take 11 weeks to acquire based on the S/N comparison given in Fig. 2 of the main text using an EO sampling trace with 500 steps and EO sampling acquisition times that we measured for our own conventional EO sampling scans. Full time-domain plots are displayed in Figs. S5 and S6.

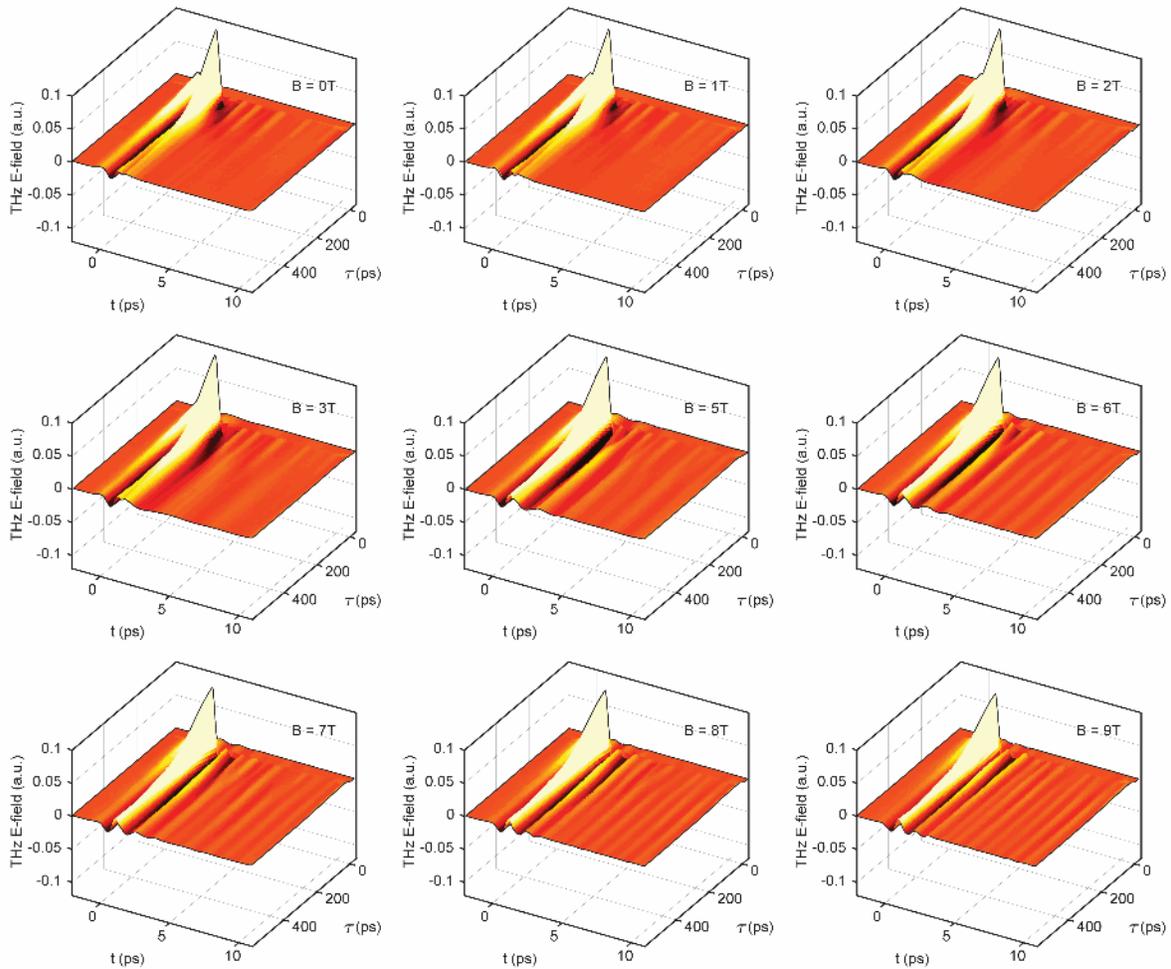

**Figure S5.** Full time-domain datasets collected at different external applied magnetic fields for the 45° sample orientation.

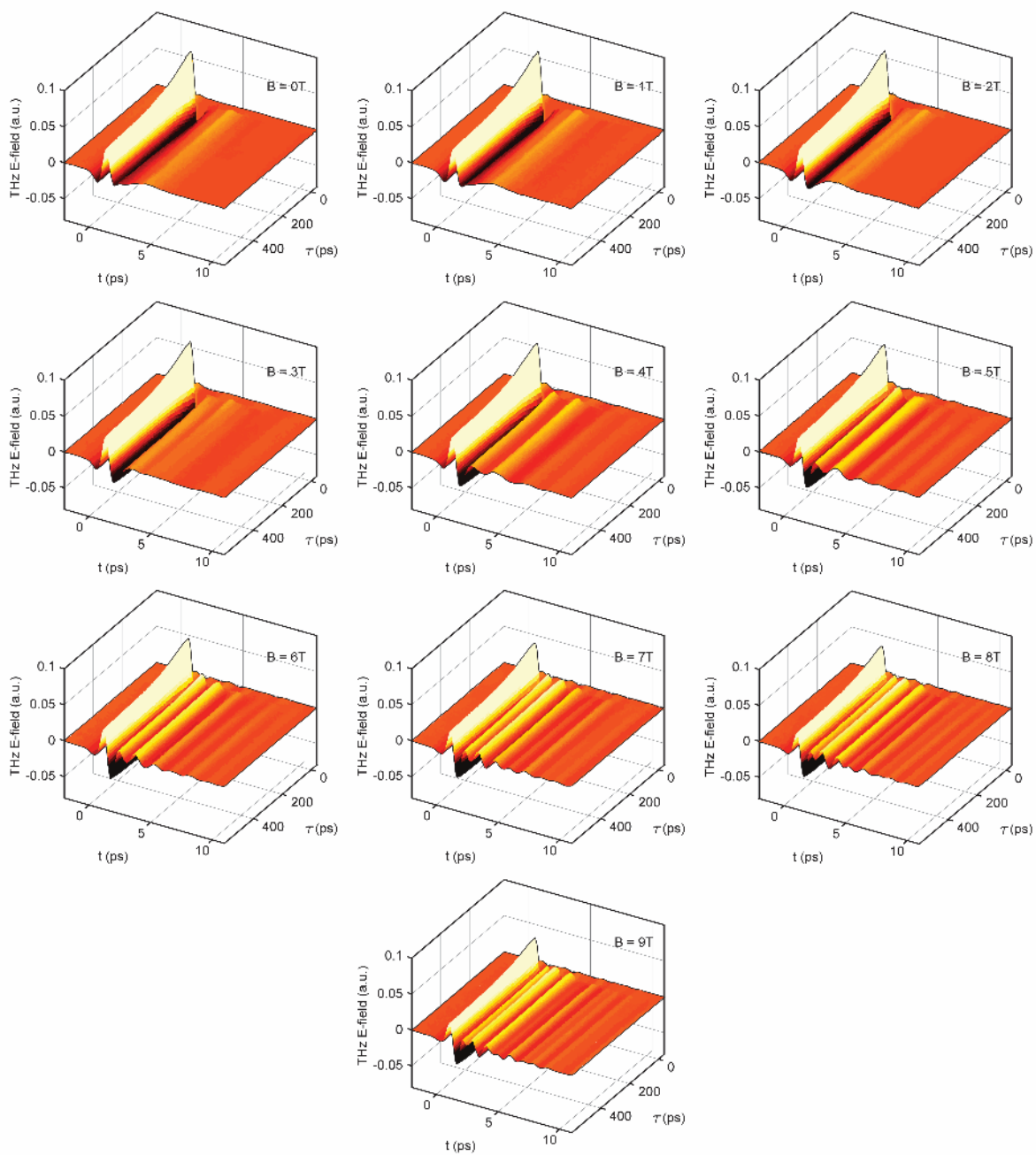

**Figure S6.** Full time-domain datasets collected at different external applied magnetic fields for the 90° sample orientation.

## S5. *Theoretical angular dependence of B-field orientation*

Theoretical expressions for the dependence of the electron and hole CR in Si on the orientation of the B-field with respect to the crystal lattice can be found in the paper by Dresselhaus et al.[1] For electrons, the analysis assumes six equivalent ellipsoidal valleys in the conduction band given by,

$$E(\boldsymbol{k}) = \hbar^2 \frac{k_\alpha^2}{m_T} + \frac{k_\beta^2}{m_T} + \frac{k_\gamma^2}{m_L} \tag{S4}$$

where $\alpha, \beta, \gamma$ are placeholders for different permutations of $x, y, z$ depending on the specific conduction band valley in question, and $m_T, m_L$ are the transverse and longitudinal effective mass parameters respectively, which for Si were given as $m_T = (0.19 \pm 0.01)m_e$ and $m_L = (0.97 \pm 0.02)m_e$. Recognizing $\hbar^2 k_{\alpha,\beta,\gamma}^2 = P_{\alpha,\beta,\gamma}^2$ where $P_{\alpha,\beta,\gamma}$ is the momentum component along $\alpha, \beta, \gamma$, and using the equation of motion, $d\boldsymbol{P}/dt = e(\nabla_{\boldsymbol{P}} E(\boldsymbol{k}) \times \boldsymbol{B})$ with $\boldsymbol{B} = B_0[\sin(\theta)/\sqrt{2}, -\sin(\theta)/\sqrt{2}, \cos(\theta)]$ where $\theta$ is the azimuthal angle of $\boldsymbol{B}$ with respect to the [100] axis in the [011] plane, we can obtain the following expressions for the electron cyclotron frequencies,

$$\begin{aligned}\omega_{[011]}^2 &= \frac{eB_0}{m_T m_L} \sin^2\theta + \frac{eB_0}{m_T^2} \cos^2\theta \\ \omega_{[100]}^2 &= \frac{eB_0}{m_L m_T} \cos^2\theta + \frac{eB_0}{2m_T^2} \sin^2\theta + \frac{eB_0}{2m_T m_L} \sin^2\theta \end{aligned} \tag{S5}$$

The effective mass is then given by $m_{[abc]}^* = eB_0/\omega_{[abc]}$, where $[abc] \in \{[100], [001]\}$.

A perturbation analysis of the spheroidal energy surfaces in the valence bands at the Γ-point yields the following expressions for the hole effective masses,

$$\begin{aligned} m_{LH}^* &= \frac{-1}{A + \sqrt{B^2 + \frac{C^2}{4}}} \cdot \left\{ 1 + \frac{C^2(1 - 3\cos^2\theta)^2}{64\sqrt{B^2 + \frac{C^2}{4}}\left(A + \sqrt{B^2 + \frac{C^2}{4}}\right)} \right\} \\ m_{HH}^* &= \frac{-1}{A - \sqrt{B^2 + \frac{C^2}{4}}} \cdot \left\{ 1 - \frac{C^2(1 - 3\cos^2\theta)^2}{64\sqrt{B^2 + \frac{C^2}{4}}\left(A - \sqrt{B^2 + \frac{C^2}{4}}\right)} \right\} \end{aligned} \tag{S5}$$

with $A = -4.1 \pm 0.2, B = 1.6 \pm 0.2$, and $C = 3.3 \pm 0.5$ as the best fit parameters for the hole CR measurements in Si conducted by Dresselhaus et al. The theoretical curves shown by the thick gray regions in Fig. S7 were obtained by a Monte Carlo simulation of eqs. S4 and S5 in which the parameters were varied within the ranges of their uncertainties.

## S6. *Effective mass uncertainty*

An upper bound on the uncertainty of the calculated effective mass values can be obtained by calculating the uncertainty at each B field from the 95% confidence intervals obtained from the Lorentzian fits. At each field, and for a given carrier type, we can calculate $\Delta m_i^* = eB_i[1/(\omega_0 - \Delta\omega) - 1/(\omega_0 + \Delta\omega)]$, where $\Delta\omega$ is half the width of the confidence interval. We can then calculate the average uncertainty $\Delta \widetilde{m}^* = \sum_i \Delta m_i^*$. Fig. S7 shows the effective masses for each carrier at the two sample orientations plotted against the theoretical curves with values and parameter uncertainties obtained from Dresselhaus et al.

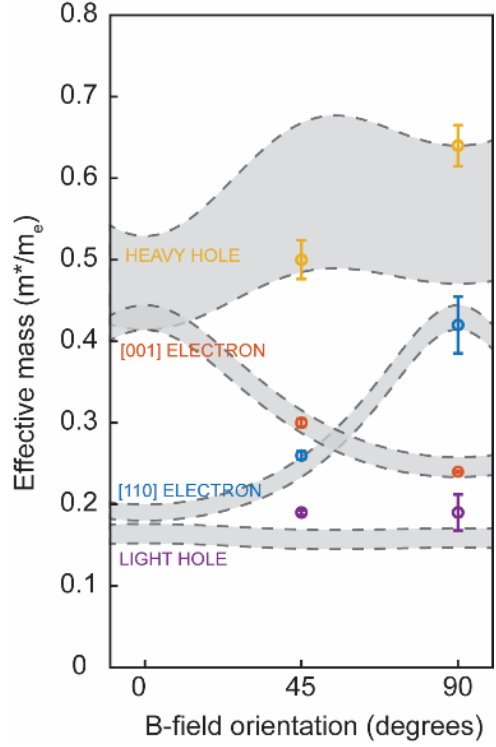

**Figure S7. Comparison of theoretical and experimental carrier effective masses.** Theoretical curves plotted in gray with bounds given by uncertainty of fit parameters from literature. Experimental values and uncertainties obtained from THz measurements as explained in the main text and section S6.

### S8. *EO sampling uncertainty due to imperfect balancing*

We can estimate the effect of imperfect balancing on an EO sampling measurement by treating the balancing imperfection as a small polarization deviation that is distributed evenly between the two detection arms. The Jones vector for the probe electric field immediately before the polarizer is then given in the basis defined by the QWP axes,

$$\boldsymbol{E}_{out} = \frac{1}{\sqrt{2}} \boldsymbol{QWP} \cdot \boldsymbol{X}_{EO} \cdot \boldsymbol{E}_{in} = \frac{1}{\sqrt{2}} \begin{bmatrix} 1+\delta \\ ie^{i\Delta\phi}(1-\delta) \end{bmatrix} \quad (S4)$$

where $\delta$ is the polarization deviation. From here we can calculate the intensity in each of the polarization arms using a polarizer at 45° with respect to the QWP axes and then calculate $\Delta I/I_0$,

$$\frac{\Delta I}{I_0} \approx \frac{I^{(+)}}{I_0^{(+)}} - \frac{I^{(-)}}{I_0^{(-)}} = \frac{1-\delta^2}{1+\delta^2} \sin(\Delta\phi) \quad (S4)$$

This is equal to the perfectly balanced EO signal multiplied by a scaling factor that depends on the size of the polarization deviation, which we define as $\Lambda \equiv (1 - \delta^2)/(1 + \delta^2)$. For deviations of ~5% as specified in the main text, we have $\Lambda = 0.995$, so that the EO sampling uncertainty for randomly distributed errors in balancing over all of the pairs of detector pixels will be on the order of $(1 - \Lambda)$, or 0.5%.

### S9. *Time dependence of cyclotron frequency and linewidth*

Figure S8 shows plots of cyclotron frequency $f_c$ at different B-field values plotted as a function of pump-probe delay. As expected, in Fig. S8 we see little variation in the frequencies of individual resonances. The variation that we do see can generally be attributed to uncertainties in the Lorentzian fitting procedure over a limited range of delays. In a few cases (e.g. $B = 7\text{T}, 90°$ orientation) the relative uncertainties are large for many delays.

Previously, work by Zhang et al. [2] established a general theory for the decay of cyclotron resonances, showing that superradiance plays a major role in decay of cyclotron resonances. They then confirmed this theory experimentally by showing a linear relationship between cyclotron resonance linewidth and carrier density as predicted by the theory. Their measurements were performed in samples with different doping levels, and hence different equilibrium carrier densities. In our measurements, we have varying carrier density as a function of pump probe delay by virtue of the carrier relaxation that occurs over the course of the 600-ps time window of our pump-probe scans. Fig. S9 shows the resonance linewidth, which we extract from Lorentzian fits of the four different resonances at each time delay, plotted as a function of pump probe delay. In some individual plots we do see modest dependence of the CR linewidth on time delay, trending in the way that we would expect (i.e. narrowing linewidth with increasing delay). This can been seen more explicity in Fig. S10, which shows the cyclotron linewidth plotted as a function of the cyclotron resonance amplitude, also extracted from the Lorentzian fitting procedure. However, in general we do not see conclusive evidence of superradiant decay, which would be manifest as a linear proportionality between CR linewidth and CR amplitude with a *y*-intercept of zero. We attribute the absence of this clear relationship to intrinsic carrier damping, which obscures any superradiant effects.

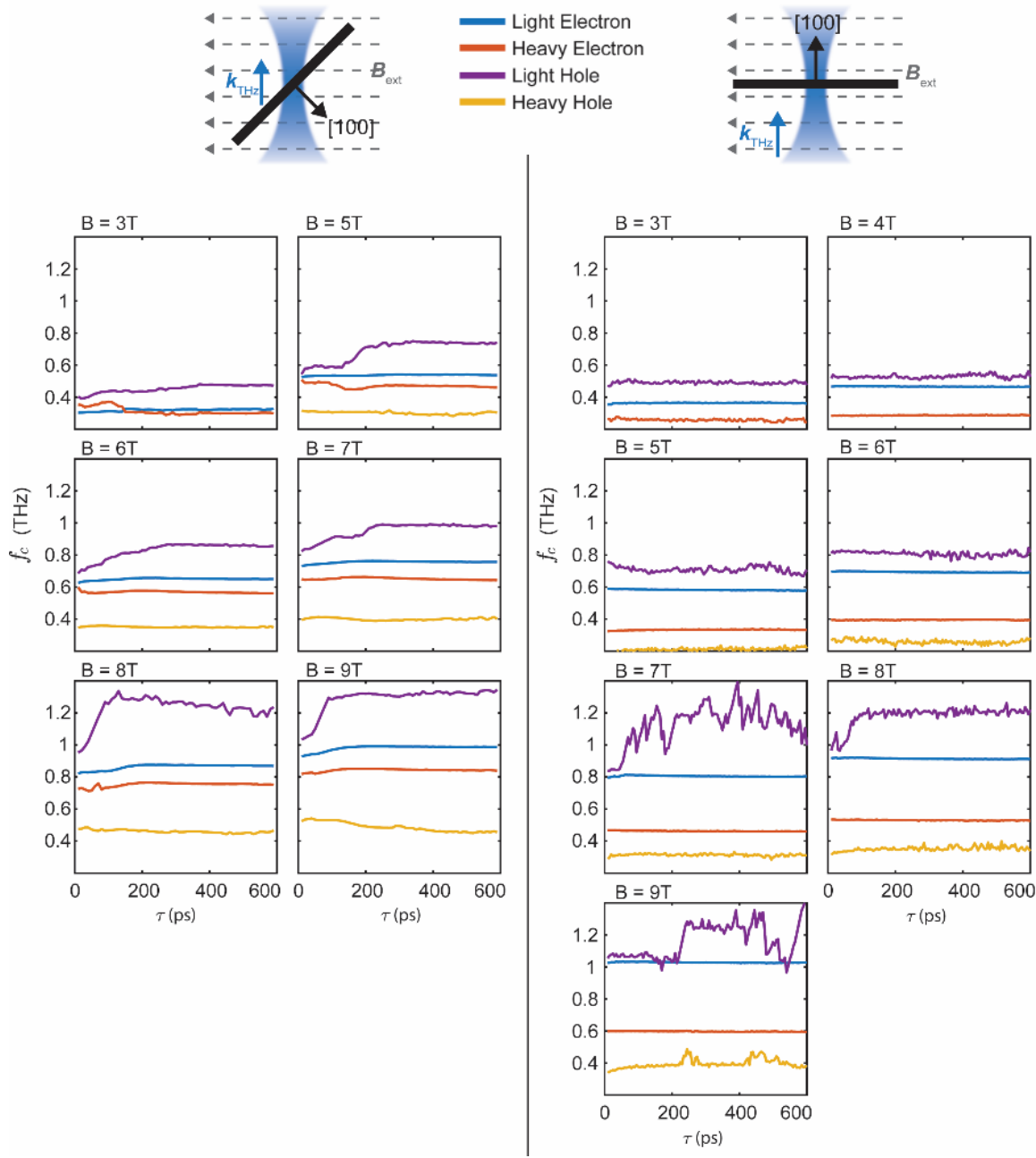

**Figure S8.** Cyclotron frequency, $f_c$, plotted as a function of pump probe delay, $\tau$.

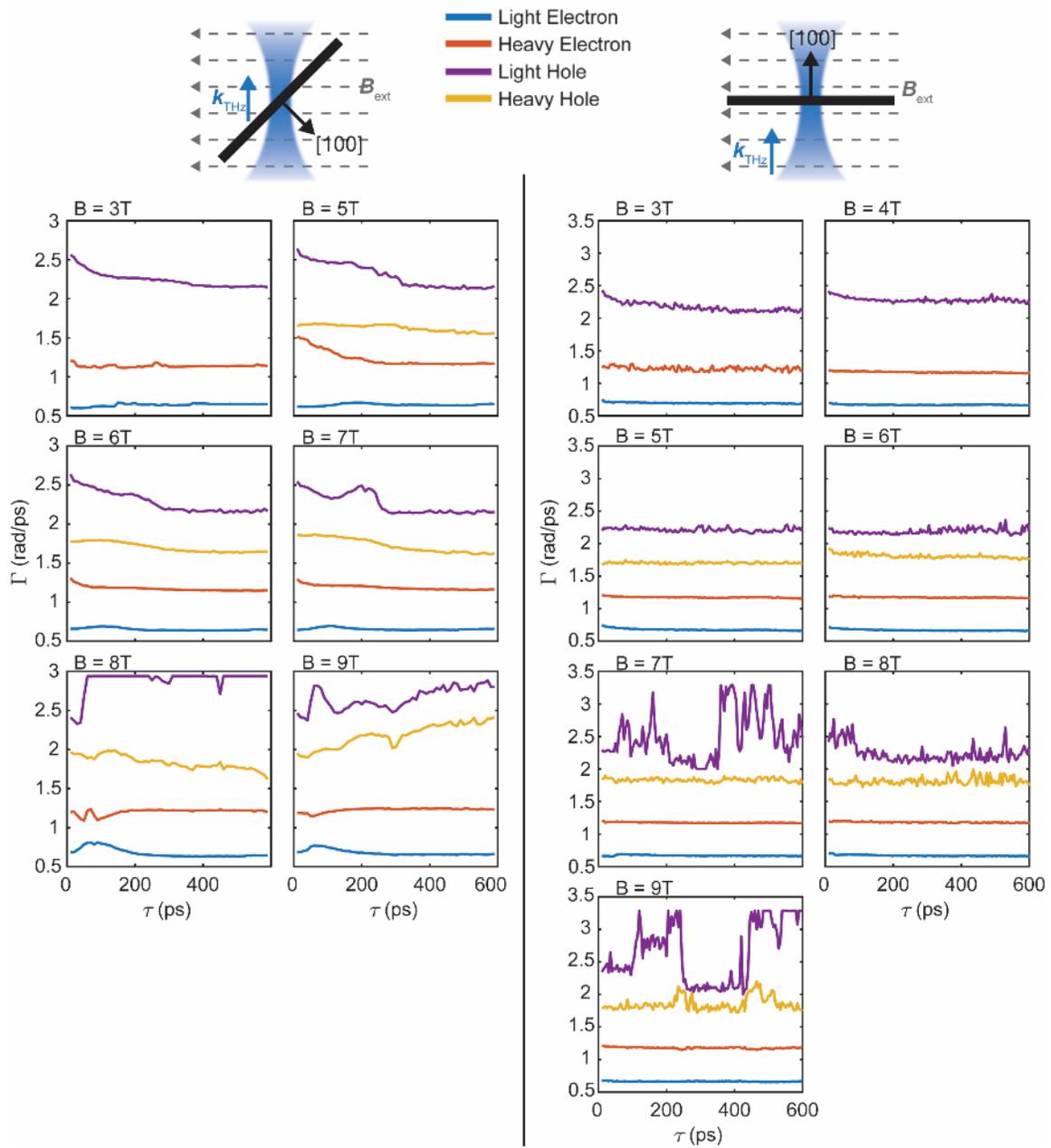

**Figure S9.** Cyclotron linewidth, $\Gamma$, plotted as a function of pump probe delay, $\tau$.

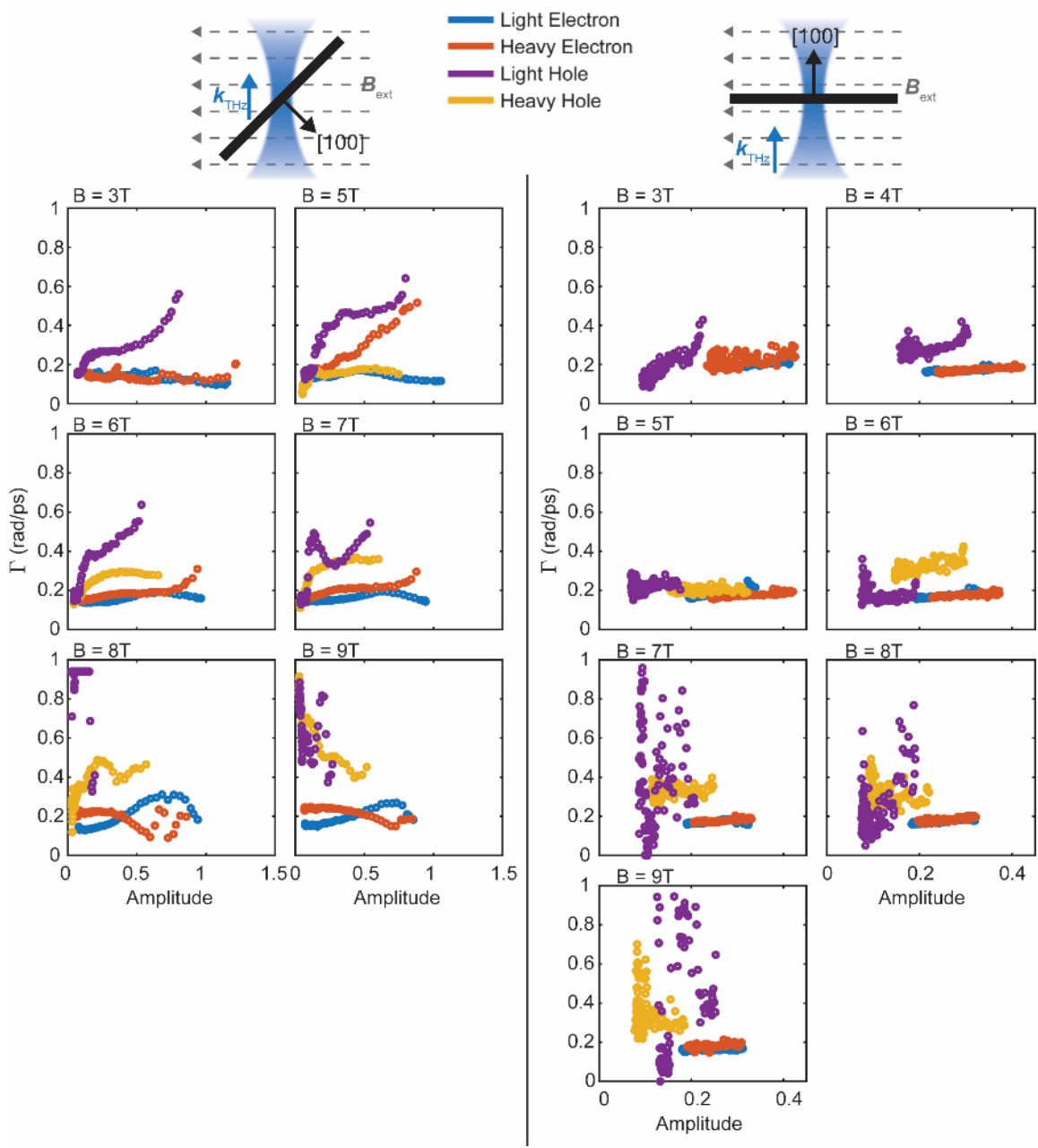

**Figure S10.** Cyclotron linewidth, $\Gamma$, plotted as a function of CR amplitude.